\documentclass{mem}
\usepackage{natbib}\usepackage{txfonts}\usepackage{balance}
\usepackage{graphicx}
\usepackage[a4paper,breaklinks,dvipdfm]{hyperref}
\idline{75}{282}
\begin{document}
\def\teff{$T\rm_{eff }$}
\def\kms{$\mathrm {km s}^{-1}$}
\def\msun{M$_{\odot}$}

\title{
Precession of White Dwarfs in CVs
}

   \subtitle{}

\author{
G. \,Tovmassian\inst{1}, S. \,Zharikov\inst{1}
\and V. \, Neustroev\inst{2}
          }

  \offprints{G. Tovmassian; gag@astrosen.unam.mx}

\institute{
Instituto de Astronomia --Universidad Nacional Autonoma
de Mexico, Mexico. 
\and
Astronomy Division, Department of Physics,  University of Oulu, Finland 
}

\authorrunning{Tovmassian }

\titlerunning{Precession of White Dwarfs}

\abstract{
Precession is observed routinely in solid bodies of Solar system and it has been invoked to explain number of phenomena observed in pulsars (i.e. Link 2003, Breton et al. 2008). White dwarfs also have been considered as possible candidates of precessing stellar objects. 
  In slowly rotating compact stars, the precession period is extremely long and the amplitude of precession is small. However, in rapid rotating neutron stars and white dwarfs, the precession period is still within reasonable observational limits and can explain observed periodicities exceeding spin periods by several times.
}
\maketitle{}

\section{Introduction}

Precession as a tool to study internal structure of stars is not new. \citet{1968ZA.....69..154M} considered the chance of detection of precession from a rotating magnetic star.  About that time  pulsars were discovered and soon thereafter \citet{1974Natur.248..483P} suggested that the observed 35 day period in Her\,X-1 may be attributed to a free precession of a neutron star. Later it was claimed that the precession of an isolated neutron star was viable and observational evidence has been found \citep{1999ApJ...524..341S,2001MNRAS.324..811J}.  Neutron stars are considered to have a solid outer crust with a liquid interior  and are expected to demonstrate free precession. Since  pulsars emit highly collimated radio beams with highly predictable timing,  the detection of precession is relatively easy. But in the case of white dwarfs, or ordinary magnetic stars which radiate isotropically, the observation of precession is practically impossible.
Binary stellar systems containing compact objects may offer the opportunity for the study of properties of matter under extreme conditions through  measurements of interactions of the compact object with the orbital companion or through derivatives of the interaction, such as accretion disks, accretion flows, etc. The obvious candidates are magnetic Cataclysmic  Variables (CVs), in which the accretion of  matter from the Roche lobe filling secondary onto the white dwarf (WD) occurs via magnetic field lines to a relatively small spot on the surface of the WD. The shock created above that spot produces a beam of highly energetic radiation observed directly in X-rays. This beam or  its reverberations from the components of the binary system can be used to detect possible precession of WD. The precession of a rapidly rotating, magnetic WD was proposed as a source of long periods detected in a few CVs, most notably FS\,Aur and V455 And  \citep{2007ApJ...655..466T}.

\begin{figure}[t!]
\resizebox{\hsize}{!}{\includegraphics[bb=45 85 710 550, clip=]{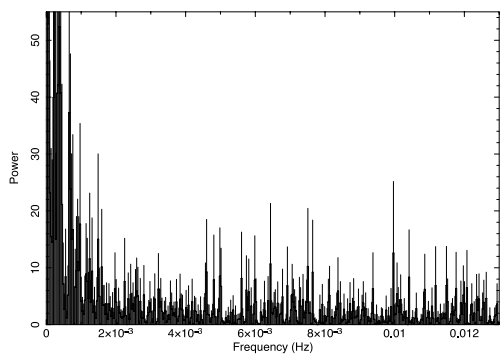}}
\caption{\footnotesize
The high frequency end of the power spectrum of the X-ray light curve of FS\,Aur
\citep{tovmassian2012}; the spin period is believed to be at 0.01 Hz, a marginal peak which nevertheless coincides with a similar result obtained from the rapid optical photometry \citep{2005MNRAS.362.1472N}.
}
\label{fsx}
\end{figure}

\section{The concept of precession and conditions to detect it}

The  idea of precession of WDs was suggested by \citet{1992A&A...261..658L}, who explicitly mention a possible application of their models to the Intermediate Polars, i.e. Cataclysmic Variables in which the magnetic WD rotation is not synchronized with the orbital period of the binary system. In this model free precession is expected to occur in a WD containing  elastic matter. Two different free precession modes are considered: that of a rigid and axially symmetric body precessing with the Euler frequency, and the Chandler frequency of an elastic body in which the instantaneous rotation axes slightly deviate from the figure axes.  
It is worth mentioning here  the fact that the elasticity  effects necessary to sustain the free precession are achieved in a carbon WDs of relatively large mass and low temperatures in which the crystallization of  the core has taken place \citep{1976ssp..conf..109V}.  In the context of CVs it means that mostly the  old, evolved systems with WDs cooled to 10\,000\,K and  mass  $\sim1$\msun\  are  candidates to contain a precessing WD.

\begin{figure}[t!]
\resizebox{\hsize}{!}{\includegraphics[angle=0, width=10cm, clip=]{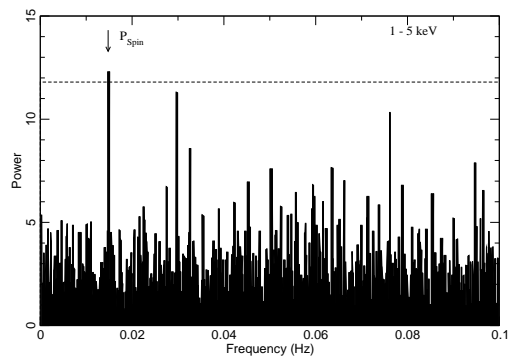}}
\caption{\footnotesize
The high frequency end of the power spectrum of the X-ray light curve of V455\,And 
\citep{gansicke2012}; the spin period is known from the  optical photometry \citep{2007ASPC..372..597G}.
}
\label{v455x}
\end{figure}%

One of the important conclusions stemming from the calculations  is that only a fast rotating WDs will have  precession rates possible to detect. \citet[see Table\,3]{1992A&A...261..658L} and \citet[see Figure\,7]{2007ApJ...655..466T}  illustrate the ratio of spin period to the precession period. The number of known systems with spin periods of order of  $10^1 - 10^2$\,s is not very large. The spin periods of IPs  predominantly cluster around   P$_{\rm {spin}} \sim0.1$\,P$_{\rm {orb}}$\  \citep{2004ApJ...614..349N}.  In 2004 only four  had  P$_{\rm {spin}}<150$ s.  Among them, only  WZ\,Sge is a short period, evolved system with possibly massive and cool WD. Since this list was published, a few discoveries have been made of short orbital period systems which show in addition long periods either photometrically \citep{2003PASP..115..725T,2011PASP..123.1156V} or spectroscopically \citep{2005A&A...430..629A,2007ApJ...655..466T}. These periods are not directly associated with the orbital or any other periods found usually in CVs. One  such object,  V455 And,  has been proven to possess a rapidly rotating and hence a magnetic WD \citep{2007ASPC..372..597G}. For GW Lib the $97\pm12$\,s spin period  estimate \citep{2010ApJ...715L.109V} is indirect, and no evidence of magnetic WD has been reported. In the case of FS\,Aur there is only  weak evidence for the  presence of the spin period  in the light curves \citep{2005MNRAS.362.1472N}. One  way to detect the spin period of the magnetic WD in a CV is by X-ray observations. FS\,Aur and V455\,And were both observed by X-ray telescopes and in both cases the signal at the presumed spin frequency is  marginal. The power spectrum of FS\,Aur  X-ray light curve is presented in Figure\,\ref{fsx}.
\begin{figure}[t!]
\resizebox{\hsize}{!}{\includegraphics[clip=true]{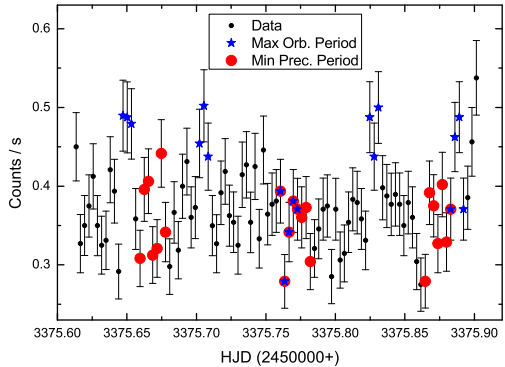}}
\caption{\footnotesize
The X-ray light curve of FS\, Aur obtained with {\sl Chandra} is presented as small dots with error bars. The stars indicate moments of X-ray pulses separated by modulo  the orbital period. The large dots mark the moments of minima in the hardness ratio modulated with the precession period. The pulse is missing when the its occurrence coincides with the hardness ratio minimum, i.e. when the pole is occulted.
}
\label{fsxlc}
\end{figure}
 Although the spin period of the WD in V455\,And is observed in the optical domain, the power in X-rays at the spin frequency is similarly low ({Figure\,\ref{v455x}).   This may sound discouraging, but in fact it is a good argument in favor of the WD precession hypothesis, because it was shown that if  the  magnetic axes and the  rotational axes of a WD are closely aligned, then there might be no  X-ray signal at the spin frequency or it will be very weak \citep{2008MNRAS.387.1157R}.   The long periods corresponding to the precession period (or the beat period between the precession and orbital periods) observed in the optical originate not on the WD itself, but in the coupling region of the accretion disk with the magnetosphere of the WD. The coupling region in magnetic CVs is known to be  highly ionized and contribute to the optical emission. It also may emit soft,  reprocessed by photo-electric absorption X-rays, and curtain the hard X-ray emission from the magnetic pole. We assume that the coupling region  is not stationary but  revolves around the inner edge of truncated accretion disk, following the magnetic pole with the precession period. And that would only happen if the magnetic pole is located at the rotational axes of the WD or very close to it.  Excess emission from that coupling region revolving with the precession period gives rise to the wings of emission lines in V455\,And and FS\,Aur as detailed in \citet{2007ApJ...655..466T}. Meanwhile, if the magnetic pole is near the equator, then regardless of whether the WD is precessing or not,  the X-ray beam sweeping around with the spin period will create an emission modulated with the spin, not precession period. The excellent example of that is demonstrated by high speed spectroscopy of DQ\,Her \citep{2010MNRAS.407.1903B,2010AJ....139.2542S}.

Thus, in order to detect precession of a WD one needs:
\begin{itemize}
\item an asynchronous  magnetic Cataclysmic variable, i.e. an Intermediate Polar.
\item the spin period of the WD in such IP must be of order of 150\,s or less.
\item the magnetic and rotation axes of the WD must be closely aligned.
\item the WD must be cool and and heavy, i.e. probable candidates should have short orbital periods.
\end{itemize}
This long list of requirement makes the detection of precession in many systems very unlikely. Also the viewing angle may weight into the probability to observe effects associated with the precession of the WD. V455\, And is a high inclination system with grazing eclipses, and the modulation in the wings of the emission lines is always clearly seen. The FS\,Aur orbital plane  has a smaller viewing angle and the modulation in emission lines was barely detected. In contrast, FS\,Aur shows prominent  photometric variability at the beat period between precession and orbital periods, something which was never observed in V455\,And. In contrast, GW Lib is a very open system, where the observer  looks perpendicular to the orbital plane and only a 4.2\,h photometric period has been detected.

\section{The modulation of the X-ray light curve with the orbital period and its relation to the precession of WD in FS Aur.} 
 
The strongest periodic signal in the X-ray light curve of FS\, Aur is the orbital. It gives a unique chance to explore the precession hypothesis. According to our model \citep{2007ApJ...655..466T} most of the X-rays, and certainly the hard emission originate from the magnetic polar region overlapping with the rotational axes. The optical emission and some of the soft X-rays  arise from the inner part of the disk. Both emitting  regions revolve with the WD with the binary period, but also independently with the precession period.  The precession may  cause an occultation of X-ray emitting region by the WD itself  with the precession period. 

After our {\sl Chandra} observations of FS\, Aur we performed two follow-up observations on this object with the {\sl Swift/XRT}  \citep{tovmassian2012}. First, the analysis of {\sl Chandra} X-ray light curve of FS\, Aur, showed that the orbital period modulation of its light curve was pulsed, i.e. it was not some sinusoidal variability but rather a series of short pulses on the otherwise flat curve.  We observed 5 orbital periods, in where we see  4 increases in flux, repeated by the orbital period if one assumes  a pulse was masked between the 2nd and 3rd pulses (Figure\,\ref{fsxlc}). If one considers separately the soft and hard X-ray light curves,  there is  a dip in the  hard X-ray flux at the time of the missing pulse, as might be expected if the  magnetic pole  partially  disappears behind the WD because of precession. We believe the disappearance of the pulse supports the hypothesis of 147\,m precession period  of the WD in a 85.7\,m orbital period  FS\,Aur.  In the Figure\,\ref{fsxlc} the orbital period phase at which the X-ray pulses appear is shown with starlike symbols and the minimums of hardness ratio modulated with the precession period are shown by large dots. The X-ray pulse is missing when its orbital phase coincides with the minimum of the precession phase. 

\begin{figure}[t]
\resizebox{\hsize}{!}{\includegraphics[clip=true]{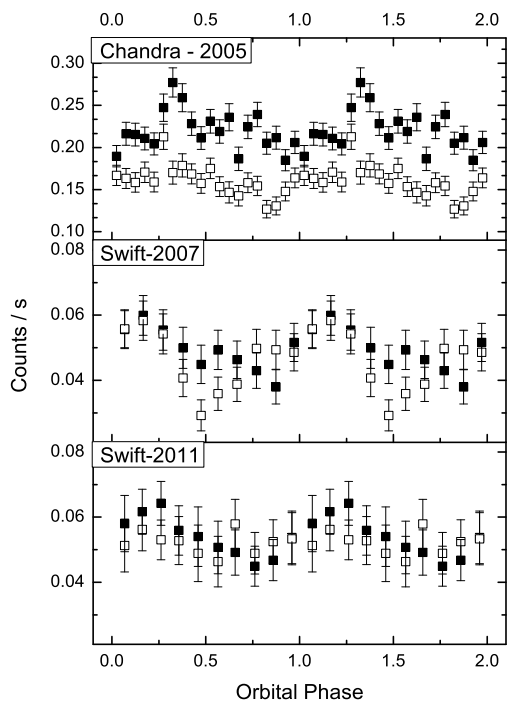}}
\caption{\footnotesize
The X-ray light curves of FS\,Aur from three separate observations folded with the orbital period.  The filled symbols are soft ($< 1.5$keV), the open symbols are hard photons ($> 1.5$keV).  The minima, more pronounced in hard energies are occurring at different orbital phases.
}
\label{fsx3y}
\end{figure}
%

Comparison of the X-ray light curve obtained by {\sl Chandra} in 2005 with the {\sl Swift} light curves in 2007 and 2011, all folded with the precise orbital period, shows that the minima in the light curves  occur at different phases from year to year
(see the presentation of Neustroev in this volume for the hardness ratio graph).  The data from {\sl Swift}   is not of high sensitivity  and the pulses and dips are not as discernible as in {\sl Chandra} observations to determine phases of  observed minima and prove that they are clocked with the precession period, but the phenomenon we have predicted is there. The  X-ray light curves folded with the orbital period  are presented in Figure\,\ref{fsx3y}. 

\section{Additional considerations.}

The sample of objects  which demonstrate the  phenomena related to the precession of the WD is very small. In addition each system is different, which in some cases provides additional support to the hypothesis, in other cases is a detriment. \citet{2010arXiv1009.5813T} show that FS\, Aur is probably a triple system.  A presence of precession might be more easily explainable in a triple system than in a simple binary. On the other hand, the long spectroscopic period of V455\,And, while present since its discovery in 2004, is not very stable. Its value varies around the averageof  3.5 hours by as much as 10\%. This is be very difficult to explain in  terms of precession model, but it must be noted, that the WD in V455\,And, as well as in GW\,Lib, is pulsating. The period of pulsations itself is not very stable. It is beyond the scope of this presentation  to discuss how  pulsations would affect the precession of a WD.

\section{Conclusions}

The show here what conditions are necessary to detect the precession of a WD in a CV. A few very strict conditions must coincide in order for us to be able to detect  precession. We also demonstrate  on the examples, that the idea of precession put forward to explain a longer than orbital optical periodicities in a number of CVs may find additional support from X-ray observations of these objects.

\begin{acknowledgements}
We are grateful to B. G{\"a}nsicke  and  E. Breedt  for providing data from papers in preparation on objects discussed in this presentation. The research on FS\,Aur and similar objects was supported financially by CONACyT grant 34521. Our research was based on X-ray observations by NASA missions {\sl Chandra} and {\sl Swift} which we acknowledge.

\end{acknowledgements}

\bibliographystyle{aa}

\end{document}